# An Empirical Study of Some Selected IR Models for Bengali Monolingual Information Retrieval

Kamal Sarkar
Computer Science & Engineering Dept.
Jadavpur University
Kolkata-700032, India
jukamal2001@yahoo.com

Avisek Gupta[*]
Electronics and Communication Sciences Unit, Indian Statistical Institute, Kolkata, India
avisek003@gmail.com

*Abstract*—This paper presents an evaluation and an analysis of some selected information retrieval models for Bengali monolingual information retrieval task. Two models, TF-IDF model and the Okapi BM25 model have been considered for our study. The developed IR models are tested on FIRE ad hoc retrieval data sets released for different years from 2008 to 2012 and the obtained results have been reported in this paper.

## I. INTRODUCTION

Information Retrieval is concerned with the retrieval of necessary information, from a large repository of information. Due to the availability of large storage spaces and high processing capabilities, it is possible to process large amounts of information. The ability to retrieve information relevant to a user's information need in limited time is the study of Information Retrieval (IR).

For a given collection of documents, an information retrieval model indexes all documents in the collection. Given a user's information need, the information retrieval model computes scores that indicate how relevant a document is to the user's information need. Documents with high relevance scores can then be retrieved.

A number of experiments have been carried out in languages such as English, where different IR models have been empirically compared to see which IR models perform the best. For Bengali, the Forum for Information Retrieval Evaluation (FIRE) [1] has conducted regular ad-hoc retrieval tracks, due to which they have gradually amassed a large collection of Bengali documents and Bengali queries. The Bengali queries have associated with it a list of relevant documents, as judged by human evaluators. Due to the FIRE ad-hoc retrieval tracks, the investigation on how various IR models performed throughout the years is available.

In this work we investigate the performance of TF-IDF and BM25 models, with an increased number of queries, to see how the models perform under a large database with a large number of queries. In the next section, we discuss the performance evaluations of various IR models that have been done at the FIRE ad-hoc retrieval task. Next, the IR models used in this work are discussed, followed by a discussion on Mean Average Precision (MAP), which is popularly used as an evaluation measure. The experiments conducted are presented next, where it is discussed how well the models have performed (1) when the number of queries is large (query set includes all queries appeared in FIRE ad-hoc retrieval tasks for the different years), and (2) when the queries are presented separately based on the year they appeared in the ad-hoc retrieval tasks.

## II. PREVIOUS WORK

Dolamic and Savoy [2] evaluated the performance of various IR models at FIRE 2008. They created a stop-words list from the corpus, and used a light stemmer that removed inflectional affixes of nouns and adjectives. They evaluated the performance of the following IR models – TF-IDF, Okapi BM25 and five models which are based on the Divergence from Randomness framework: PB2, GL2, PL2, I($n_e$)B2 and I($n_e$)C2. Here each of models is denoted by XYZ having components X, Y and Z where X is the name of a model of randomness, Y is the first normalization method and Z is the second normalization method. For example, if the name of IR model is PB2, which means X is "P" (that is, model of randomness is Poisson), first normalization is "B" (it is one of two normalization schemes) and second normalization is "2" (one of two normalization schemes). The highest MAP they achieved was 0.4131 from the I($n_e$)B2 model.

McNamee [3] in FIRE 2008 evaluated four language models where the corpus was indexed using unigram, 4-gram, 5-gram, and 4-grams that skipped a single interior letter. The highest MAP for Bengali was 0.3582 for the 5-gram language model.

Paik and Parui [4] developed a stemmer where words having common prefix were defined to be in the same class, and the words were replaced by their common prefix. Using the IR model IFB2 (a variant of "divergence from randomness" model [2]) with a prefix length 3, they achieved maximum MAP of 0.4232 in the official run of FIRE 2008.

TABLE 1. SUMMARY OF PREVIOUS WORK ON FIRE AD HOC RETRIEVAL

| Authors | Method | Year of FIRE ad hoc retrieval task | Evaluation (MAP) |
|---|---|---|---|
| Dolamic, Savoy[2] | TF-IDF, BM25, language model, PB2, GL2, PL2, I($n_e$)B2 and I($n_e$)C2 | FIRE 2008 | 0.4131 |
| McNamee[3] | unigram, 4-gram, 5-gram, and 4-grams skipped | FIRE 2008 | 0.3582 |
| Paik and Parui[4] | Common prefix based stemmer, IFB2 | FIRE 2008 | 0.4232 |
| Loponen et al[5] | YASS, StaLe, GRALE stemmers | FIRE 2010 | 0.5190 |
| Bhaskar et al[6] | TF-IDF, position and distribution factors, theme-based document clustering | FIRE 2010 | 0.4002 |
| Leveling et al[7] | Term Conflation, language modeling, BM25 with Blind Relevance Feedback | FIRE 2010 | 0.341 |
| Dolamic, Savoy[8] | TF-IDF, BM25, language model, PB2, GL2, PL2, I($n_e$)B2 and I($n_e$)C2 | FIRE 2010 | 0.5026 |
| Paik et al[10] | Frequent Case Generation | FIRE 2011 | 0.3457 |
| Banerjee and Pal[9] | Frequency-based stemmer | FIRE 2011 | 0.3182 |
| Ganguly et al[11] | Decompounding : Relaxed and Selective | FIRE 2008 FIRE 2010 FIRE 2011 FIRE 2012 | 0.3148 0.4352 0.3279 0.2985 |
| Barman et al[12] | Query expansion using Wikipedia, Entropy-based ranking | FIRE 2012 | 0.0438 |

Loponen et al [5] compared the performance of the YASS stemmer with StaLe and GRALE lemmatizers for FIRE 2010. Using YASS yielded the best MAP of 0.5190 when evaluating using the Indri search engine of the Lemur Project[1].

Bhaskar et al [6] for FIRE 2010 used a stop-word removal and suffix stripping module on the corpus. For each term a combination of TF-IDF, position and distribution factors were used to assign a weight to it. Top *n* words were identified as keywords, representing some theme.

All documents in the corpus were clustered based on these keywords so that documents with the same keywords representing similar themes would lie in the same cluster. Given a query, the closest documents were retrieved using cosine distance. Their method received a MAP score of 0.4002.

Leveling et al [7] evaluated Term Conflation at FIRE 2010. For Term Conflation they used three approaches: reducing words to *n* prefixes, a corpus-based stemmer and a rule-based stemmer. They used language modeling to index the corpus, and used Okapi BM25 with Blind Relevance Feedback to evaluate their method. Among the three Term Conflation methods, the best MAP result they obtained was 0.341 for a 5-prefix reduction.

Dolamic and Savoy [8] at FIRE 2010 followed methods similar to the methods employed at FIRE 2008. The same stop-words list and light stemmer were applied to the corpus. They evaluated the following IR models – TF-IDF, Okapi Bm25, Language Modeling and four Divergence from Randomness framework models – PB2, GL2, PL2 and I($n_e$)C2. For Bengali they achieved best MAP results of 0.5026 for the I($n_e$)C2 model.

Banerjee and Pal [9] developed a stemmer for FIRE 2011. The frequency-based stemmer they developed showed performances that were comparable to YASS. Using YASS, they achieved a MAP of 0.3435, and using their frequency-based stemmer, they achieved a MAP of 0.3182.

For FIRE 2011, Paik et al [10] evaluated their method Frequent Case Generation, which is a fast alternative to lemmatization where for each word a number of different forms are generated by statistical analysis of the corpus. For Bengali they achieved a maximum MAP of 0.3457 which was comparable to the MAP of the n-gram model, which was 0.3501.

Ganguly et al [11] investigated the effect of decompounding for Bengali IR. They found out that the standard approach of decompounding did not work very well. Instead, they proposed two approaches, relaxed decompounding and selective decompounding. They obtained MAP values of 0.3148 for FIRE 2008 queries, 0.4352 for FIRE 2010 queries, 0.3279 for FIRE 2011 queries, and 0.2985 for FIRE 2012 queries.

Barman et al [12] for FIRE 2012 performed Query expansion using Wikipedia and performed Entropy-based ranking. They achieved a maximum MAP of 0.0438.

A summary of previous work on FIRE ad hoc retrieval is presented in table 1.

III. DESCRIPTION OF DATASET

The Forum for Information Retrieval Evaluation (FIRE) for their 2012 ad hoc retrieval task compiled a dataset of Bengali documents. The dataset contains 500122 documents. It contains news articles from reputed newspapers Anandabazar Patrika and BDNews. From Anadabazar Patrika it contains articles from the year 2001 to the year 2010. From BDNews, it contains articles from the year 2006 to the year 2010.

---
[1] www.lemurproject.org

All files are encoded in UTF-8. Our indexed collection has a total of 500121 documents. The queries that were used in various FIRE ad hoc retrieval tasks, and the number of documents in the corpus of that year, are shown in the table 2.

During FIRE 2008 and 2010, a smaller subset of the dataset was used for the ad hoc retrieval task. Therefore, the queries for those years have human relevance judgments only on the smaller subset of documents that existed in the dataset during that time. More specifically, the queries 26 to 125 have human relevance judgments for news articles of Ananadabazar Patrika from September of 2004 to September of 2007, and no documents from BDNews. For queries 126 to 225, human relevance judgments exist over all documents in the dataset.

TABLE2. NUMBER OF BENGALI QUERIES IN FIRE AD HOC RETRIEVAL TASKS FOR DIFFERENT YEARS

| Queries | Number of documents | FIRE ad hoc retrieval task year |
|---|---|---|
| 26 to 75 | 123021 | 2008 |
| 76 to 125 | 123021 | 2010 |
| 126 to 175 | 500121 | 2011 |
| 176 to 225 | 500121 | 2012 |

In our experiments we have used two datasets. One dataset is the subset of the dataset used for human relevance judgment of queries 26 to 125. That dataset is used to evaluate queries 26 to 125. And the other dataset is the entire dataset, used to evaluate queries 126 to 225.

## IV. INFORMATION RETRIEVAL MODELS

The primary aim of Information Retrieval is to build an Information Retrieval (IR) model that analyses each document and extract the necessary information from it and assigns a score to each document in response to a given query. The documents are then produced in a list, ordered in decreasing order of the score assigned to them by the IR model, where the document most relevant to the query is at the top of the list. The IR models used in the experiment are discussed next, after discussing the concept of the Bag-of-Words model, which is a general way of extracting information from documents.

### A. The Bag-of-Words Model

The Bag-of-Words model views a document as a collection of words. The only information that is collected from a document is what words are present in it. It rejects all syntactic information from a document.

Given a collection of documents $C$, containing words from a vocabulary $V$, the following information can be extracted from each document.

*Term Frequency (TF):* For a word $w_i$, the Term Frequency $TF(w_i, d_j)$ measures the frequency of $w_i$ in document $d_j \in C$. If $w_i$ is not present in $d_j$, $TF(w_i, d_j) = 0$. Usually $TF(w_i, d_j)$ is simply written in a shortened form $TF$ when it is present in more complex formulas.

*Document Frequency (DF):* For a word $w_i$, the $DF(w_i)$ measures the number of documents in the collection $C$ the word $w_i$ is present in. $DF$ is used to calculate the Inverse Document Frequency $IDF$, which is an important measure in IR.

*Inverse Document Frequency (IDF):* $IDF(w_i)$ is the inverse of the $DF$. So if a word $w_i$ is rare, it has a low $DF$, and its $IDF$ is high, and if $w_i$ is present in a large number of documents in the collection, it has a high $DF$, and its $IDF$ is low. IDF is calculated using the formula:

$$IDF(w_i) = \log\left[\frac{N+1}{DF(w_i)+1}\right]$$

Where: N is the number of documents in a collection.

*Document Length:* The Document Length (*DocLen*) of a document $d_j$ is the number of words contained in it.

$$DocLen(d_i) = \sum_{w_i \in d_i} TF(w_i, d_i)$$

*Average Document Length:* This is the average over all documents in the collection. If a collection $C$ contains $n$ documents, then the Average Document Length (*AvgDocLen*) is computed as shown below

$$AvgDocLen = \frac{1}{n} \sum_{d_i \in C} DocLen(d_i)$$

### B. The TF-IDF Model

Given a query and a collection of documents, a retrieval model should retrieve the documents most relevant to the query, Q. The TF-IDF model combines the *TF* and *IDF* of query words in a document leading to a retrieval that is better than using *TF* or *IDF* alone [13]. This is justified by the fact that a higher frequency of query words present in a document should indicate that the document is more relevant to the query. Thus higher *TF* values should contribute to higher scores for a document. Also, query words that are rare in the collection should be more capable of distinguishing between documents that are relevant to the query, compared to words that are present in many documents. So, higher *IDF* values should contribute to higher scores for a document.

Using the Lemur Project [14], we implemented a score function for the basic TF-IDF model. The TF-IDF score for a document d is:

$$\text{TF-IDFScore}(d) = \sum_{w_i \in Q \land w_i \in d} TF(w_i, d) \times IDF(w_i)$$

A small difference in the *TF* or *IDF* values can lead to a large change in the score. Therefore we also implemented a popular variant of the basic TF-IDF score, the log TF-IDF score, where the logarithm of both *TF* and *IDF* are taken.

Some difficult situations are faced if the logarithm of *TF* and *IDF* is used. When *TF* is 0, the calculation of logarithm of 0 will be required. To avoid this situation, the logarithm of (1 + *TF*) is calculated. *IDF* value is calculated as follows:

$$IDF(w_i) = \log\left(\frac{(1+N)}{(1+DF(w_i))}\right)$$

To avoid situation where a word is unseen (*DF* = 0), *DF*+1 is used instead of *DF*. (*N*+1) is used instead of *N*. So the score for the log TF-IDF score is given as follows.

$$\log TF\text{-}IDFScore(d) = \sum_{w_i \in Q \wedge w_i \in d} \log(1+TF(w_i,d)) \times \log\left(\frac{(1+N)}{(1+DF(w_i))}\right)$$

### C. The Okapi BM25 model

Okapi BM25 [15] [16] is a probabilistic retrieval model which has the following score function.

$$BM25Score(d,q) = \frac{TF}{TF + (K1*(1-B)) + B \times (DLen/AvgDLen)} \times IDF$$

Where d is a document and q is the query.
It is similar to the TF-IDF function, having a *TF* part and an *IDF* part in its score function [17]. The *TF* part has certain restrictions imposed on it that lead to better scoring of documents. The basic TF-IDF score is linearly proportional to *TF*. Large *TF* values can have a dominating effect on a document's score. To normalize *TF*, a parameter K1 is used, to change *TF* to $\frac{TF}{TF + K1}$. The parameter B is used to normalize document lengths, resulting in the score of the BM25 model.

BM25, as implemented by the Lemur Project, assigns a score to each document given a query term by the following score function.

$$LemurBM25Score(d,q) = \frac{K3+1}{K3} \times IDF \times TFfactor$$

$$\text{where} \quad TFfactor = \frac{(1+K1) \times TF}{TF + (K1*(1-B)) + B \times (DocLen/AvgDocLen)}$$

Here there is a slight change in the way K1 is used to set an upper bound to *TF*. K3 is a parameter that adds weight to the entire score.

### V. EVALUATION

The evaluation metric Mean Average Precision (MAP) [18] is used to evaluate the IR models. The MAP evaluation metric requires for each query, the list of ranked documents that is output by a retrieval model, and the list of documents relevant to a query as evaluated by a user. Let the set of queries be $Q = \{q_1, q_2, ..., q_j\}$. If for a query $q_j$ there are $m_j$ relevant documents in the output list of ranked documents, then average precision (*AP*) is calculated as follows.

$$AP(q_j) = \frac{1}{m_j} \sum P(k)$$

Here *k* is the position of a relevant document in the ranked list, and *P(k)* is the precision at position *k*. Precision at position *k* is calculated as follows.

$$P(k) = \frac{rel\_k}{k}$$

where *rel_k* is the number of relevant documents retrieved till the position *k*. MAP is calculated by computing the average of *AP* over all queries.

$$MAP(Q) = \frac{1}{|Q|} \sum AP(q_j) = \frac{1}{|Q|} \sum \frac{1}{m_j} \sum P(k)$$

Therefore using MAP, a single numeric metric is obtained that can be used to evaluate a retrieval model.

In our experiments, we have used *trec_eval* [19], a module for evaluating relevance judgment of IR systems. It accepts as input two files, a query-relevance file, and a results file. The query-relevance file contains the human relevance judgment of each query, in the format shown below.

> *query-number 0 document-id relevance*

Here *query-number* is the number of a query, which is followed by a constant *0*. *document-id* is the ID representing a document, and *relevance* can have two values : 0, indicating the document with the *document-id* is not relevant to the query with the given *query-number*, or 1, indicating the document is relevant to the query.

The results file contains the ranked list of documents returned by the IR system. The format of results file is shown below.

> *query-number Q0 document-id rank score Exp*

Here *query-number* is the number of a query, *Q0* and *Exp* are constants used by certain evaluation software, *document-id*

is the ID of a retrieved document, *rank* is the document's position in the ranked list, *score* is the score assigned to the document by the IR system.

The *trec_eval* module returns several information based on the ranked list and human evaluation. Mean Average Precision (MAP) is returned, along with several other measures such as Average Precision Geometric Mean, R-Precision, Precision after a certain number of documents (5,10,15,...) were retrieved, etc. Among these measures, MAP is popularly used to evaluate IR systems, due to which we have used it to evaluate the IR systems we investigated.

## VI. EXPERIMENTS AND RESULTS

The Lemur Project was used to build an inverted index on the two datasets as discussed in section 3, and score functions were incorporated in it to develop the IR models which are evaluated on the sets of queries. The dataset was pre-processed before it was indexed using Lemur. First punctuation was removed from all documents. Stop-words were then removed using the list of stop-words provided by FIRE. Stemming was performed next using the Yet Another Suffix Stripper (YASS) [20]. Documents containing stemmed words and no punctuation were written in the TREC format (a format for NIST's Text REtrieval Conference). The same removal of punctuation and stemming was performed on the queries. Lemur indexed these two datasets, and the indices were used to evaluate the different IR models. The score function for TF-IDF model and log TF-IDF model were implemented, and the implementation of the score function for BM25 that exists in Project Lemur was used.

In contrast with the TF-IDF and log TF-IDF models, the BM25 score has 3 parameters, K1, B and K3, which need to be tuned for obtaining the better retrieval performance.

TABLE 3. MAP FOR DIFFERENT SET OF QUERIES FOR THE IR MODELS

| Model | MAP for queries 26 to 125 | MAP for queries 126 to 225 |
|---|---|---|
| TF-IDF | 0.2451 | 0.0668 |
| log TF-IDF | 0.3326 | 0.1878 |
| BM25 | 0.4733 | 0.3518 |

For tuning the parameters of our developed BM25 model, it was evaluated for values of K1 ranging from 0.2 to 3.0, in increments of 0.2. The parameter B was changed from 0 to 1, in increments of 0.2. It was observed that for high values of K3, BM25 usually performed better. So K3 was iterated with the values 20, 100, 150, 200, and 300. The MAP evaluation for all models is shown in the table 3. For BM25, the best MAP values obtained are shown.

For queries 26 to 125, BM25 model with K1 set to 2.2, B set to 0.4, and K3 set to 100 gave the best score of 0.4733. For queries 126 to 225, BM25 model with K1 set to 1.0, B set to 0.6, and K3 set to 20 gave the best score of 0.4733.

Our developed BM25 model was also run on year-wise query sets. After using the YASS stemmer, the maximum MAP values achieved using BM25 model for each set of queries is shown in the table 4, along with the corresponding parameter values for K1, B, and K3 of the BM25 model.

TABLE 4. MAP FOR YEAR-WISE SET OF QUERIES

| Year | Queries | K1 | B | K3 | MAP |
|---|---|---|---|---|---|
| 2008 | 26-75 | 2.0 | 0.6 | 250 | 0.4177 |
| 2010 | 76-125 | 2.2 | 0.2 | 70 | 0.5313 |
| 2011 | 126-175 | 1.0 | 0.6 | 20 | 0.3608 |
| 2012 | 176-225 | 0.8 | 0.4 | 20 | 0.3539 |

### A. Comparisons to the Existing Systems

For system comparisons, we have chosen the systems which had been tested on the same FIRE ad hoc retrieval data sets and the results are reported in the literature.

Table 1 presented in section 2 summarizes the performances of the various IR models tested on FIRE ad hoc retrieval data sets released in the different years. Comparing table 1 and table 4, we have observed that the MAP values obtained by our developed systems were at par with, or sometimes even better, than the MAP values achieved by other participants of the FIRE ad hoc retrieval task. For some cases, where the basic model used by a FIRE participant is similar to our developed model, we observe that we obtain better results in comparison to their models. The main reason may be the proper tuning of the system parameters (for example, OKAPI BM 25 model, there are a number of parameters which need to be properly tuned for better results) and the variations in stemming algorithm used in the model.

## VII. CONCLUSION

In our work, we have designed a simple IR system which includes removal of punctuation and stop-words, stemming using YASS, and ranking using BM25. This IR system has performance comparable to other systems investigated in the IR literature. Some alternatives to the YASS stemmer or the BM25 IR model can be employed to see how well they perform.

The BM25 or TF-IDF models works on documents, treating documents as bag-of-words. It will be interesting to see how adding the ability to analyze syntactic information to the IR system changes its retrieval performance. IR models can be enhanced with adding the ability to recognize synonyms and/or recognizing similar phrases.

ACKNOWLEDGMENT

This research work has received support from the project entitled ''Design and Development of a System for Querying, Clustering and Summarization for Bengali'' funded by the Department of Science and Technology, Government of India under the SERB scheme.

---

[*] Avisek Gupta took part in carrying out the work presented in this paper when he worked as a Junior Research Fellow in the SERB (DST) funded project entitled ''Design and Development of a System for Querying, Clustering and Summarization for Bengali" under Department of Computer Science Engineering, Jadavpur University.